\documentclass[aps,preprint]{revtex4}

\usepackage{graphicx}
\usepackage{amsmath, amsthm}
\begin{document}

\title{{\itshape Ordering effects in diluted magnetic semiconductors}
}

\author{Josef Kudrnovsk\'y$^{1}$, V. Drchal$^{1}$,Georges Bouzerar$^{2}$ and Richard Bouzerar$^{3}$}

\affiliation{
$^{1}$ Institute of Physics AS CR,CZ-182 21 Prague, Czech Republic\\
$^{2}$ Institut N\'eel CNRS and Institut Laue \& Langevin,
F-38 042 Grenoble, France\\
$^{3}$ Universit\'e de Picardie Jules Verne,
F-80 039 Amiens, France
}

\begin{abstract}
We review recently developed two-step approach for 
description of electronic and magnetic properties of 
a new class of materials, the diluted magnetic 
semiconductors.
In the first step we construct, on the basis of the 
state-of-the-art first-principles electronic structure 
calculations, the effective Ising and Heisenberg Hamiltonians 
which describe, respectively, the alloy phase stability and 
the magnetic excitations in the system.
In the second step, we analyze properties of these effective
Hamiltonians by various methods of statistical physics.
As a case study, the prototypical diluted magnetic semiconductor
Ga$_{1-x}$Mn$_{x}$As is studied in detail.
We determine, among others, a possibility for clustering in this
system, formation energies of various compensating defects, and
estimate short-range order parameters describing ordering
tendencies in a system.
On the other hand, by using recently developed local random-phase 
approximation approach, we evaluate the system Curie temperature
and demonstrate its strong dependence on the sample preparation.
We also emphasize the relevance of proper inclusion of the 
randomness in positions of magnetic impurities for a reliable 
estimate of the system critical temperature.
Finally, we compare calculated Curie temperatures with available 
experimental data and briefly mention relation to other theoretical 
approaches.
\end{abstract}

\maketitle

\section{Introduction}

Properties of semiconductors are very sensitive to a small amount
of various impurities and defects while magnetism is a collective
phenomenon often stable to high temperatures.
Magnetic order in the system strongly influences other material
properties, like e.g., the phase stability, and transport and
optical properties.
Also, unique properties of both semiconductors and magnetic materials
form basis of many important technologies.
The combination of properties of both these classes of materials
seems to be very promising and in fact resulted recently \cite{Ohno}
into the discovery of a new class of materials, the diluted magnetic
semiconductors (DMS), of which the diluted III-V DMS is the best
example and (Ga,Mn)As system is the most frequently studied one
both experimentally and theoretically.

In addition to being promising in future spintronics applications,
the DMS represent also the challenge to the solid state and material 
science physics.
A main feature of these new materials is ferromagnetism which is
primarily due to coupling of impurity magnetic moments mediated
by free carriers, typically by holes, in the host semiconductor
valence band.
These holes are introduced into the GaAs host by Mn-impurities 
but their concentration in the system can be strongly influenced 
by the presence of native defects, namely by As-antisites on the
Ga-sublattice and by Mn-interstitials.
The hybridization between magnetic impurity and host electronic
states leads to a formation of virtual bound states for which
proper inclusion of electron correlations can be quite important.
However, one of the most important features characterizing the 
DMS is the presence of the disorder and, in particular, how the
disorder influences the distribution of magnetic and other 
impurities: the atom clustering or segregation, the tendency of 
foreign atoms to be incorporated into the system in the presence of 
other impurities, both native and doping ones, etc. are relevant
questions to be addressed.
In addition to these basic structural issues there is an equally
important question of how important can be disorder for the Curie
temperature of the system: sufficiently robust ferromagnetism is
a key point for any future technological application of the DMS.

There is an extensive literature on the theory of DMS 
\cite{rev1,rev2,database} which, however, summarizes mostly results 
of model approach to the problem.
The parameter-free, first-principles studies are less frequent 
and a systematic, comprehensive review on this subject is still 
missing.
The aim of this paper is not to give such a review but rather to
illustrate one specific approach which employs the first
principles calculations as a starting point for the construction
of simple, effective Hamiltonians which could address various
important problems from the fields of the phase stability and
magnetic excitations in a simple, transparent way rather than 
to attempt to study these problems fully on the first-principle
level, which is still in most cases numerically prohibitive.  
Below we will consider two such models, the Ising and Heisenberg 
Hamiltonians, whose parameters, the effective pair interactions
in the former case, and the pair exchange interactions in latter 
case, are determined on the basis of the same first principles 
calculations.
In both cases such parameters can be determined either from 
{\it ad hoc} chosen structural or magnetic configurations in
the framework of conventional, electronic structure supercell 
calculations \cite{ising,philmag} or, as in the present case, 
by explicitly including the effect of disorder in terms of the 
coherent-potential approximation (CPA) \cite{book}.
One obvious advantage of the present approach is the
possibility to include the effect of small concentrations 
of various types of impurities as well as the 
effect of finite lifetime due to disorder, features typical
for the DMS, on the same footing.
On the other hand, the effect of clustering on the electronic
and magnetic structure can be more straightforwardly included 
using the supercell approach.
A detailed comparison of both approaches is, however, beyond 
the scope of the present paper. 
An emphasis will be put on the study of magnetic properties of GaMnAs
alloys, in particular on the determination of their Curie temperatures
based on a detailed study of the corresponding Heisenberg Hamiltonian
by sophisticated statistical methods which take into account both random
positions of Mn-impurities and the presence of compensating defects in
a system.

\section{Electronic structure}
\label{ESM}

We have determined the electronic structure of the DMS in the framework
of the first principles all-electron tight-binding linear muffin-tin
orbital (TB-LMTO) method in the atomic-sphere approximation using empty 
spheres in interstitial tetrahedral positions of the zinc-blende lattice 
for a good space filling.
We used equal Wigner-Seitz radii for all atoms and empty spheres.
The valence basis consists of $s$-, $p$-, and $d$-orbitals, we include
scalar-relativistic corrections but neglect the spin-orbit effects.
The substitutional disorder due to Mn-atoms and other possible defects 
is included within the CPA.
The charge selfconsistency is treated in the framework of the local
spin density approximation using Vosko-Wilk-Nusair parameterization
for the exchange-correlation potential \cite{VWN}.
The lattice constant of the pure GaAs ($a=5.653~{\rm \AA}$) was used
in all calculations but we have verified that we can neglect a weak
dependence of the sample volume on defect concentrations.
Further details of the method can be found in \cite{book}.

The  magnetic disorder is treated in the framework of the disordered 
local moment (DLM) method \cite{dlm} which is the simplest
way of including disorder in spin orientations and which is
justified for atoms with large exchange splitting.
A comparison of the total energies of the ferromagnetic (FM) state
with its DLM counterpart is the simplest way to investigate the
magnetic stability of the DMS alloy \cite{satomfa,our}.
In addition, the DLM is a natural starting point for the phase 
stability studies due to the fact that sample preparation is done 
at temperatures well above the system Curie temperature.
The DLM can be included in the framework of the CPA: the Mn atoms
have collinear but random positive (Mn$^{+}$) and negative (Mn$^{-}$)
orientations with corresponding concentrations $x^{+}$ and $x^{-}$,
$x=x^{+}+x^{-}$, where $x$ is the total Mn-concentration.
The degree of magnetic order can then be characterized by the order
parameter $r=(x^{+}-x^{-})/x$, and $x^{\pm}=(1 \pm r)x/2$.
In the (FM) state, $r=1$, all magnetic moments are aligned in the
direction of a global magnetization.
The non-magnetic state, $r=0$, is characterized by a complete
disorder of spin directions with vanishing total magnetization while
a partial ferromagnetic state is characterized by $0~<~r~<~1$.
For more details concerning the determination of the magnetic phase
diagram and magnetic moments in different magnetic states we refer
to \cite{mpd,mpd2}.

\begin{figure}
\centering
\includegraphics[width=0.75\textwidth]{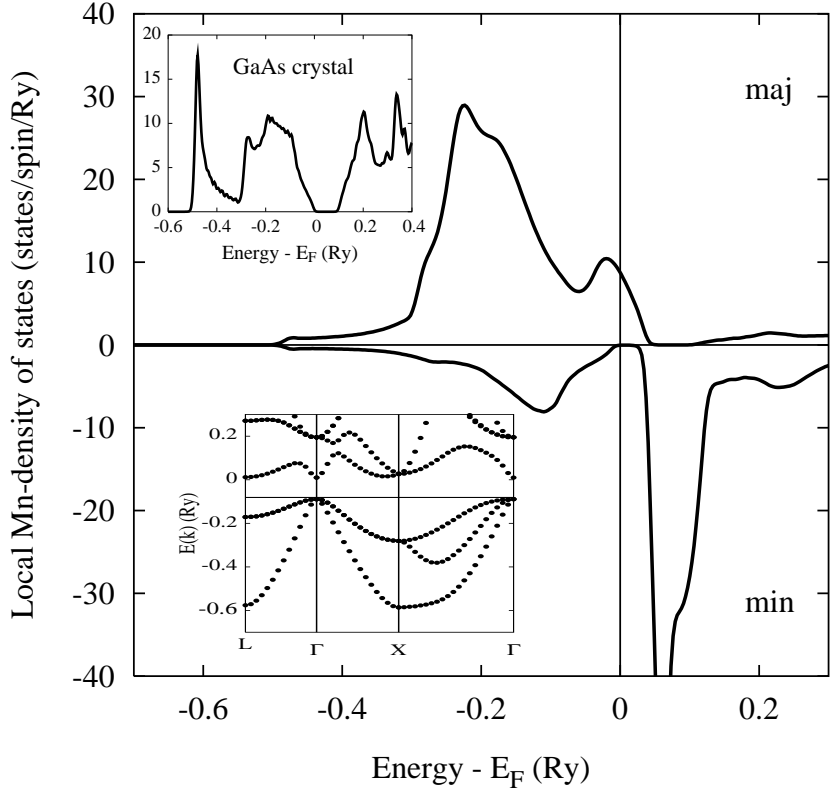}
\caption{Electronic structure of (Ga$_{0.05}$,Mn$_{0.05}$)As random
alloy. The upper and lower insets show the total density of states 
and the bandstructure along principal axis in the Brillouin zone
of the host GaAs crystal.
The spin-polarized local density of states on Mn-atoms substituting
cations on Ga-sublattice is shown in the main frame.}
\label{Fig.1}
\end{figure}

In Fig.~\ref{Fig.1} we plot the local density of states on Mn-atoms,
clearly illustrating general features of the GaMnAs alloys which
will be relevant for their magnetic properties: (i) the halfmetallic 
character of spin-subbands with a gap in the minority states, and
(ii) the strong disorder introduced by the presence of Mn-impurities 
in the GaAs host as manifested by its pronounced difference from 
the host GaAs density of states.
Also the reference host bandstructure is shown for an illustration.

\section{Ising Hamiltonian of an alloy}

Some important problems concerning the structure and phase 
stability of DMS can be studied on the basis of the Ising 
Hamiltonian describing various configurations of disordered 
alloys with parameters determined from first principles.
A particular configuration of a homogeneous disordered
multicomponent alloy is characterized by occupation indices
$\eta_{\bf R}^{Q}$, where $\eta_{\bf R}^{Q} = 1$ if
the site ${\bf R}$ is occupied by an atom of type $Q$,
and $\eta_{\bf R}^{Q} = 0$ otherwise.
Configurational averaging of occupation indices
$\langle \eta_{\bf R}^{Q} \rangle = c^{Q}$
yields the concentrations $c^{Q}$.
The configurationally dependent part of the alloy internal
energy is given by the effective Ising Hamiltonian
\begin{equation}
H = + \frac{1}{2} \sum_{\bf R R'} \sum_{ Q Q'} V_{\bf R R'}^{Q Q'} \,
\eta_{\bf R}^{Q} \, \eta_{\bf R'}^{Q'}
+ \dots \, ,
\label{ising1}
\end{equation}
where $V_{\bf R R'}^{QQ'}$ are interatomic pair interactions. 
Here we will not consider higher order interactions.

The pair interactions $V_{\bf R R'}^{Q Q'}$ in DMSs consist
of two contributions,
\begin{equation}
V_{\bf R R'}^{Q Q'}=v_{\bf R R'}^{Q Q'}+\phi_{\bf R R'}^{Q Q'}
 \, ,
\label{vrr}
\end{equation}
where the $v_{\bf R R'}^{Q Q'}$ result from mapping of the band
part of the total energy onto the Ising Hamiltonian (\ref{ising1})
and the $\phi_{\bf R R'}^{Q Q'}$ represent the electrostatic
interaction energy of a pair of atoms $Q,Q'$ located at sites 
${\bf R, R'}$ (for derivation see Appendix~A in \cite{VDPhilMag04})
$\phi^{QQ'}_{\bf R R'} =
e^2 q^{Q}_{\rm eff} q^{Q'}_{\rm eff}/|{\bf R - R'}|$,
where $q^Q_{\rm eff}=q^Q-\bar{q}$ is the effective net charge of 
atomic species $Q$ defined as a difference of the net charge $q^Q$ 
of atomic species $Q$ and the averaged charge $\bar{q}$.
The band term contribution is calculated using the Generalized
Perturbation Method (GPM)\cite{ising,VDGPM96,book}
\begin{equation}
v^{QQ'}_{\bf R R'} =
 \frac{1}{\pi} {\rm Im}\int^{E_{\rm F}}_{E_{\rm min}} {\rm d} E \,
{\rm tr} \, \Bigl[  t_{\bf R}^{Q}(z) \bar{g}_{\bf R R'}(z)
t_{\bf R'}^{Q'}(z) \bar{g}_{\bf R' R}(z) \Bigr] \, ,
\label{bandv}
\end{equation}
where tr denotes a trace over angular momentum indices $(\ell m)$
and the spin index $\sigma$, $z=E+i0$, $E_{\rm F}$ is the CPA Fermi
energy, $E_{\rm min}$ is a suitably chosen energy below the valence
energy spectrum, $\bar{g}_{\bf R R'}(z)$ denotes the block of the
averaged auxiliary Green function between sites ${\bf R}$ and
${\bf R'}$, and $t_{\bf R}^Q(z)$ is the t-matrix for atomic
species $Q$.

\subsection{Impurity formation energies}

The calculated total energies (per elementary cell) make possible to
investigate impurity formation energies for DMSs.
The formation energy $\varepsilon[{\rm A}_{\rm B}]$ of an impurity
A$_{\rm B}$ which substitutes a host atom B in a (generally multicomponent)
alloy A$_x$B$_{1-x}$ is defined as
\begin{eqnarray}
\varepsilon[{\rm A}_{\rm B}] =
N E[{\rm A}_{x+\delta x}{\rm B}_{1-x-\delta x}] + E_{\rm at}[B]
- \{ N E[{\rm A}_{x}{\rm B}_{1-x}] + E_{\rm at}[A]  \} \, ,
\label{ife1}
\end{eqnarray}
where $N$ is the number of elementary cells in the alloy,
$\delta x = 1/N$, $E_{\rm at}[{\rm A}]$ is the energy of an
isolated atom A, and $E[{\rm A}_{x}{\rm B}_{1-x}]$ is the energy
of the alloy per one elementary cell.
By expanding into linear terms in $\delta x$ one finds
\cite{Masek02}
\begin{equation}
\varepsilon[{\rm A}_{\rm B}] =
 \frac{\partial E[{\rm A}_{x}{\rm B}_{1-x}]}{\partial x} +
E_{\rm at}[B] -  E_{\rm at}[A] \, .
\label{ife2}
\end{equation}

Our calculations \cite{VDPhilMag04} for (Ga$_{1-x-y}$Mn$_{x}$As$_{y}$)As 
alloys have shown that the formation energy of As antisite defect
decreases with increasing concentration $x$ of substitutional Mn atoms,
which means that the number of the antisite defects can be considerably 
enhanced in the presence of substitutional Mn.
Similarly, the formation energy of the substitutional Mn
decreases with an increasing concentration $y$ of As antisites.
This means that the presence of As antisites (and probably also of
other donors \cite{Masek02}) is important for an improved solubility 
of Mn in III-V materials.

The behavior of formation energies for interstitial Mn impurities 
and for As antisites in the presence of interstitials is opposite 
to that found for substitutional Mn because increasing concentration 
of one of the species leads to a growth of the impurity formation 
energy of the other one.
This is in agreement with the growth mechanism of
(Ga$_{1-x-y}$Mn$_{x}$As$_{y}$)As alloys discussed by Erwin and
Petukhov \cite{Erwin}: the Mn atoms are first incorporated into interstitial
positions under low concentration of As antisites, and later on, during
growth or annealing, as substitutional impurities.
In particular, the conversion of the interstitial Mn into a substitutional
form is facilitated during the growth process if a sufficient
number of As antisites is available.
Both discussed mechanisms contribute to the self-compensation behavior 
of as-grown (Ga,Mn)As alloys.

\subsection{Alloy stability }

The calculated total energies (per elementary cell) make possible to
investigate the stability of DMSs with respect to segregation 
into systems with different chemical composition.
For example, consider segregation of (Ga$_{1-x-y}$Mn$_{x}$As$_{y}$)As 
into an alloy without As-antisites (Ga$_{1-x}$Mn$_{x}$)As and an alloy 
with the highest possible concentration of As-antisites which is still 
not overcompensated, (Ga$_{1-3x/2}$Mn$_{x}$As$_{x/2}$)As,
\begin{eqnarray}
\Delta E(x,y) &=& E[({\rm Ga}_{1-x-y} {\rm Mn}_{x} {\rm As}_{y}){\rm As}]
         -\frac{x-2y}{x} E[({\rm Ga}_{1-x}{\rm Mn}_{x}){\rm As}]
\nonumber \\
         && -\frac{2y}{x} E[({\rm Ga}_{1-3x/2}{\rm Mn}_{x}{\rm As}_{x/2}){\rm As}] \, .
\label{segr1}
\end{eqnarray}
The energy $\Delta E(x,y)$ is negative \cite{VDPhilMag04} which indicates 
the stabilizing effect of the As antisites.
This is in agreement with the above conclusions based on the impurity
formation energies.

Similarly, we can consider segregation of the alloy (Ga$_{1-x}$Mn$_{x}$)As 
into pure GaAs and an alloy with higher concentration of Mn atoms, say $x_0$,
(Ga$_{1-x_0}$Mn$_{x_0}$)As,
\begin{eqnarray}
\Delta E(x,x_0) &=& E[({\rm Ga}_{1-x} {\rm Mn}_{x} ){\rm As}]
  -\frac{x_0-x}{x_0} E[{\rm Ga As}]
\nonumber \\
 && \!\!\!\!\!\!\!\! -\frac{x}{x_0} E[({\rm Ga}_{1-x_0}{\rm Mn}_{x_0}){\rm As}] \, .
\label{segr2}
\end{eqnarray}
Our calculations \cite{VDPhilMag04} have shown that this energy is positive
which shows that the alloy (Ga$_{1-x}$Mn$_{x}$)As is thermodynamically 
unstable with respect to segregation.

\subsection{\label{TORD} Ordering tendencies}

The effective interatomic pair interactions $V_{\bf R R'}^{Q Q'}$ 
in semiconductors decrease rather slowly with the interatomic distance 
$|{\bf R - R'}|$ and have to be calculated over many coordination spheres 
\cite{VDPhilMag04}.
Also their Coulombic part, even though weak, is long-ranged.
In order to analyze possible ordering patterns, we employed the linearized 
concentration wave method and for temperatures above the ordering temperature 
$T_{\rm ord}$ we also calculated Warren-Cowley short-range order parameters.
One has to keep in mind that in these methods only the configurational 
part of the entropy is taken into account which usually leads to an 
overestimation of ordering temperatures.

The ordering temperature $T_{\rm ord}$ and the type of ordered
structure that appears below $T_{\rm ord}$ can be studied in terms
of the concentration-wave method \cite{Khachat,ising}.
Here we employ its linearized version \cite{Bose97} extended to a
multicomponent alloy since we consider possible ordering of four atomic
species (Ga, Mn$^\uparrow$,  Mn$^\downarrow$, As) on the cation
sublattice.
In a mean field approximation (i.e., assuming a Bragg-Williams
form of the entropy \cite{Huang}) the free energy is expressed
in terms of local concentrations $c^Q_{\bf R}$,
\begin{equation}
F =  \frac{1}{2} \sum_{{\bf R R'}} \sum_{ Q Q'}
 V_{\bf R R'}^{QQ'} c_{\bf R}^Q c_{\bf R'}^{Q'} +
k_{\rm B} T \sum_{{\bf R}} \sum_{Q} c_{\bf R}^Q \, {\rm ln}(c_{\bf R}^Q)
 \, ,
\label{free1}
\end{equation}
where $k_{\rm B}$ is the Boltzmann constant and $T$ is the temperature.
Starting from the disordered state, the free energy can be expanded
up to quadratic terms in concentration fluctuations
$\delta c_{\bf R}^Q = c_{\bf R}^Q - c^Q$,
\begin{equation}
F = F_0 + \frac{1}{2} \sum_{{\bf R R'}} \sum_{ Q Q'}
\Bigl[ V_{\bf R R'}^{QQ'} + \frac{k_{\rm B} T}{c^Q}\delta_{\bf R R'}
\delta_{QQ'} \Bigr] \,
\delta c_{\bf R}^Q \, \delta c_{\bf R'}^{Q'}
 \,
\label{free2}
\end{equation}
with terms linear in $\delta c_{\bf R}^Q$ vanishing because
$\sum_{\bf R} \delta c_{\bf R}^Q = 0$ for all $Q$, and
$\sum_{{\bf R'}Q'} V_{\bf R R'}^{QQ'} c^{Q'}$ is a constant
for all ${\bf R}$ and $Q$ (for details see Ref. \cite{Bose97}).
Here $F_0$ is the free energy in the absence of concentration
waves ($c_{\bf R}^Q = c^Q$ for all ${\bf R}$).
Equation (\ref{free2}) can be rewritten in terms of a lattice
Fourier transform in a matrix notation as
\begin{equation}
\Delta F = F - F_0 = \frac{1}{2} \sum_{\bf k}^{\rm BZ} Y^{\dagger}({\bf k})
\Bigl[ V({\bf k}) + k_{\rm B} T C^{-1} \Bigr]
Y({\bf k})  |\epsilon({\bf k})|^2
 \, ,
\label{free4}
\end{equation}
where $[V({\bf k})]_{QQ'} = V^{QQ'}({\bf k})$, and the matrix $C$
is defined as $[C]_{QQ'}=c^Q \delta_{QQ'}$.
In Eq.~(\ref{free4}) the concentration fluctuations
$\delta c^{Q}({\bf k})$ are expressed in terms
of a vector $Y({\bf k})$ and the order parameter $\epsilon({\bf k})$ as
$\delta c^{Q}({\bf k}) = Y^Q({\bf k}) \epsilon({\bf k})$.
At sufficiently high temperatures $\Delta F$ is positive definite,
because the hermitian matrix $V({\bf k})+k_{\rm B} T C^{-1}$ has
only positive eigenvalues and thus the high temperature state is
completely disordered ($\epsilon({\bf k})=0$ for all ${\bf k}$).
With decreasing temperature it can become indefinite
at $T_{\rm ord}$ because of a vanishing eigenvalue for a critical
vector ${\bf k}_0$ which determines the period of the
concentration wave.
The components $Y^Q({\bf k})$ of the critical eigenvector determine
the amplitude of the concentration wave for each alloy component $Q$.
For each ${\bf k}$, the minimization of $\Delta F$, and thus
the eigenvalue problem, is subject to the subsidiary condition
$\sum_Q Y^Q({\bf k})= 0$ which follows from
$\sum_Q \delta c^Q_{\bf R}= 0$, valid for each ${\bf R}$.
The ordering temperature is then found as the largest
eigenvalue of the matrix
${\Theta}({\bf k}) = - k_{\rm B}^{-1} C^{\frac{1}{2}}
V({\bf k}) C^{\frac{1}{2}}$.

For example, for a ferromagnetic alloy
(Ga$_{0.93}$Mn$^{\uparrow}_{0.06}$As$_{0.01}$)As
we found \cite{VDPhilMag04} the ordering temperature
$T_{\rm ord} = 775$~K and the ordering vector
${\bf k}_0 = 0.274 (1,1,1)/a$, where $a$ is the lattice constant.
A closer examination shows that the largest eigenvalues of
the matrix $\Theta({\bf k})$ have very similar values
(within 1 K) for ${\bf k}$-vectors close to a surface
of a sphere of radius $0.479/a$, which corresponds
to a domain of a characteristic radius 3.7 nm in real space.
The components of the eigenvector $Y({\bf k}_0)$ are
$Y({\rm Ga})= 0.726$, $Y({\rm Mn}^\uparrow)= -0.686$,
and $Y({\rm As})= -0.040$, which corresponds to formation
of domains of two types:
in the first type the concentration of impurities Mn and As
is increased, while in the second type the impurity
concentrations are diminished in agreement with the above
estimate based on the segregation energy (\ref{segr2}).

\subsection{\label{APPSRO} Short-range order parameters}

The Warren-Cowley short-range order parameters \cite{ising}
\begin{equation}
\alpha_{\bf R R'}^{QQ'} =
- \frac{\langle \eta_{\bf R}^{Q} \eta_{\bf R'}^{Q'} \rangle
- \langle \eta_{\bf R}^{Q}\rangle
  \langle \eta_{\bf R'}^{Q'} \rangle}
{\langle \eta_{\bf R}^{Q}\rangle \langle \eta_{\bf R'}^{Q'} \rangle}
= 1 - \frac{\langle \eta_{\bf R}^{Q} \eta_{\bf R'}^{Q'} \rangle}
      {c^{Q} c^{Q'}}
\label{defal}
\end{equation}
provide detailed information on mutual correlations of
impurities; they also can directly be used in calculations
of transport and magnetic properties.
The matrix of the Warren-Cowley parameters can be approximately
calculated by means of the Krivoglaz-Clapp-Moss (KCM) formula
in reciprocal space,
\begin{equation}
\alpha({\bf k}) =
-D [M+\beta \widetilde{V}({\bf k})]^{-1} D^T \, ,
\label{KCMfor}
\end{equation}
where  $\beta = (k_{\rm B} T)^{-1}$, and the matrix $M$ is
defined as
$[M]_{Q Q'} = 1/c^0 + \delta_{Q,Q'}/c^Q$.
The matrix $D$ is introduced to ensure a correct normalization
of $\alpha$ for ${\bf R}={\bf R'}$ which follows from the definition
in (\ref{defal}).
Note that $\widetilde{V}({\bf k})$ is a Fourier transform of 
$\widetilde{V}_{\bf R R'}^{QQ'} = V_{{\bf R R'}}^{QQ'}
+ V_{{\bf R R'}}^{00} - V_{{\bf R R'}}^{Q0} - V_{{\bf R R'}}^{0Q'}$ ,   
which is the matrix of pair interactions from which one atomic species
was eliminated.
The inverse lattice Fourier transform of (\ref{KCMfor}) then yields 
the Warren-Cowley parameters in the real space.

The results for the Warren-Cowley parameters show a strong tendency 
to an aggregation of Mn atoms with the same orientation of spin.
This finding is in line with the results of Ref. \cite{Schilfg}.
Close pairs of Mn atoms with opposite moments become possible,
while the probability to find close pairs of As antisites is
very low.
A moderate aggregation of Mn atoms and As antisites is also
possible.

The results for the Warren-Cowley parameters show a strong tendency
to an aggregation of Mn atoms with the same orientation of spin.
This finding is in line with the results of Ref. \cite{Schilfg}.
Close pairs of Mn atoms with opposite moments become possible,
while the probability to find close pairs of As antisites is
very low.
A moderate aggregation of Mn atoms and As antisites is also
possible.

\subsection{\label{CONCL} Conclusions from structural studies}

The main conclusions can be summarized as follows:\\
(i) The alloys are thermodynamically unstable with respect to
segregation into related compounds or alloys.\\
(ii) As-antisites have a stabilizing effect and make the
incorporation of substitutional Mn atoms energetically favorable.
On the other hand, incorporation of Mn atoms into interstitial
positions is energetically favorable only at low concentration
of As antisites.\\
(iii) Formation of domains of two types, namely, with an enhanced
and with a lowered concentration of impurities (substitutional Mn
atoms and As antisites), can be expected.
The characteristic size of the domains depends on chemical composition
and might be of order of several nm.\\
(iv) A strong tendency to aggregation of substitutional Mn atoms 
with parallel magnetic moments is expected, while the formation of 
close pairs of As antisites is highly unlikely.\\

\section{Heisenberg Hamiltonian}

The knowledge of exchange interactions allows one to address
in detail the character of magnetic excitations in the DMS, i.e.,
to evaluate the Curie temperature, the spin-wave stiffness, and the
spectrum of low-lying magnetic excitations.
Magnetic excitations in ferromagnets are of two different kinds, namely,
Stoner excitations associated with longitudinal fluctuations of the
magnetization, and spin-waves, or magnons corresponding to collective
transverse fluctuations of the magnetization direction.
The low-temperature regime is dominated by magnons and Stoner 
excitations can be usually neglected.  
In this section we will construct an effective random Heisenberg 
Hamiltonian with classical spins which will be used in the next
section to study the critical temperatures.
The main idea is to separate a relevant part of (very) small
magnetic energies responsible for magnetic ordering from
huge total energies obtained from first-principles
total energy calculations \cite{lie}.
The validity of this approach, based on the adiabatic approximation,
is in particular justified for magnetic atoms with large exchange
splitting, like, e.g., Mn-impurities in GaAs host.
The mapping is further simplified by using the magnetic force-theorem
\cite{lie,ozb} which states that the band energy of the calculated
ground state of a reference spin structure can be used as an estimate
for corresponding total-energy differences in the excited state.
Note that intracell non-collinearity of the spin polarization is
neglected since in this approach we are primarily interested in
low-energy excitations due to intercell non-collinearity.
The application of this approach to disordered systems like the DMS 
is significantly simplified by using the vertex-cancellation theorem 
(VCT) which justifies the neglect of disorder-induced vertex 
corrections in Eq.~(\ref{eq_Jij}) below.
The VCT was derived in \cite{vct} under rather general conditions 
and facilitates an efficient evaluation of exchange interactions, 
exchange stiffnesses, spin-wave energies, etc.
The present real-space approach is particularly suitable for random
systems with low concentrations of magnetic impurities such as the DMS,
where the effect of disorder is treated in the framework of the CPA.
It is also possible to estimate exchange interactions from energy
differences between parallel and various antiparallel configurations 
of a few magnetic atoms in a supercell representing specific 
concentration (the real-space supercell approach) \cite{schm}.
In this way exchange interactions of magnetic clusters could be 
estimated approximately.
Alternatively, the reciprocal-space approach to the problem is also
possible, namely the frozen-magnon method combined with the supercell
approach \cite{sb}.
We refer reader to a recent review \cite{philmag} for more details,
comparison of the real-space and reciprocal-space approaches, 
as well as for applications of the above formalism to a broad range 
of various magnetic systems.

\subsection{\label{EI} Effective pair exchange interactions}

The mapping of the total energy of itinerant electron system 
connected with small rigid rotations of two magnetic moments at 
sites $\bf R$ and $\bf R'$ can be described by the effective 
Heisenberg Hamiltonian with classical spins as discussed above
\begin{equation}
H_{\mbox{\scriptsize eff}} = - \sum_{{\bf R} \neq {\bf R'}}
J_{\bf R,R'} {\bf e}_{\bf R} \, \cdot {\bf e}_{\bf R'} \, .
\label{eq_HH}
\end{equation}
Here, $J_{\bf R,R'}$ is the exchange interaction energy between 
sites $({\bf R,R'})$, and ${\bf e}_{\bf R}, {\bf e}_{\bf R'}$ are unit 
vectors pointing in directions of local magnetic moments at sites 
$({\bf R,R'})$, respectively.
In the present formulation the values and signs of the magnetic moments
are already absorbed in the definition of the $J_{\bf R,R'}$'s so that 
positive (negative) $J_{\bf R,R'}$'s correspond to ferromagnetic 
(antiferromagnetic) coupling.
By adopting the magnetic force-theorem \cite{lie,ozb,philmag}, the 
configurationally averaged effective pair exchange interactions 
${\bar J}^{M,M'}_{\bf R,R'}$ between two magnetic atoms $M,M'$
located randomly at sites $\bf R$ and $\bf R'$ are given by the
following expression \cite{our}:
\begin{equation}
{\bar J}^{M,M'}_{\bf R,R'} = \frac{1}{4\pi} \, {\rm Im} \int_{C} \,
 {\rm tr}_L \, \Big[ \delta^{M}_{\bf R}(z) \,
{\bar g}^{M,M' \uparrow}_{\bf R,R'}(z) \,
\delta^{M'}_{\bf R'}(z) \, {\bar g}^{M',M \downarrow}_{\bf R',R}(z) \Big]
\, {\rm d} z \, .
\label{eq_Jij}
\end{equation}
Here, ${\rm tr}_{L}$ denotes the trace over angular momenta 
$L=(\ell m)$, the energy integration is performed in the upper 
half of the complex energy plane along a contour $C$ starting below 
the bottom of the valence band and ending at the Fermi energy, and
$\delta^{M}_{\bf R}(z)= 
P_{\bf R}^{M,\uparrow}(z)-P_{\bf R}^{M,\downarrow}(z)$, where the
$P_{\bf R}^{M,\sigma }(z)$ are the $L$-diagonal matrices of
potential functions of the TB-LMTO method for
$\sigma=\uparrow ,\downarrow $ corresponding to a particular magnetic
atom $M$.
The matrix $\delta^{M}_{\bf R}(z)$ reflects the exchange splitting
of atom $M$.
The quantities ${\bar{g}^{M,M' \uparrow }}_{\bf R,R'}(z)$ and
${\bar{g}^{M',M \downarrow }}_{\bf R',R}(z)$ refer to site off-diagonal
blocks of the conditionally averaged Green function \cite{book},
namely, the average of the Green function over all configurations
with atoms of the types $M$ and $M'$ fixed at sites $\bf R$ and 
$\bf R'$, respectively, determined in the framework of the CPA 
\cite{book}.
The exchange interactions between magnetic moments induced on 
non-magnetic atoms are negligible as compared to the exchange 
interactions between magnetic atoms in the present case, i.e., 
$M$=$M'$=Mn in (Ga,Mn)As alloys.
The main advantage of the present approach is the explicit expression 
for ${\bar J}^{M,M}_{\bf R,R'}$ which can be evaluated straightforwardly 
even for large distances $d=|{\bf R}-{\bf R}|$ between sites $\bf R$ and 
$\bf R'$ and thus allowing the study of their asymptotic behavior as a 
function of the interatomic distance $d$ \cite{our,philmag}.
It should be noted that the rigidity of spins during rotations as
required by the Heisenberg model is preserved by construction in the
present approach and in the reciprocal-space method \cite{sb} but 
not in the real-space supercell approach \cite{schm}. 

The effect of impurities on the host bandstructure is usually 
neglected in model theories \cite{rev1,rev2}, but it need not be 
a small perturbation in the presence of the virtual-bound states 
\cite{levy}. 
It is also relevant for DMS systems as it was demonstrated recently 
\cite{GBRBTZ} by using the selfconsistent local RPA theory \cite{EPL}:
the combined effect of random geometry and thermal fluctuations is
crucial and calls for a proper treatment.
In particular, it was shown that damped RKKY interactions often used 
in model studies, are unable to represent properly the effect of 
virtual bound states in the host band on the values of exchange 
interactions, and corrections are needed (see also \cite{tiam}).
The standard RKKY exchange interactions cannot explain ferromagnetism
observed in the DMS: one has to go further in this perturbation
scheme to include properly effects of resonances due to virtual
bound states.
On the other hand as already mentioned, in the present approach is
the effect of virtual bound states included by construction.

The neglect of the effect of impurities on exchange interactions
means that the unperturbed host Green function appears in 
(\ref{eq_Jij}) rather than its conditionally averaged counterpart 
${\bar{g}^{{M,M'} \sigma}}_{\bf R,R'}(z)$.
The neglect of the renormalization of the host Green function by 
scatterings on impurities is two-fold: it introduces a phase factor 
and modifies the amplitude of the oscillations as compared to 
the conventional RKKY formula \cite{rkky}.

\begin{figure}
\centering
\includegraphics[width=0.75\textwidth]{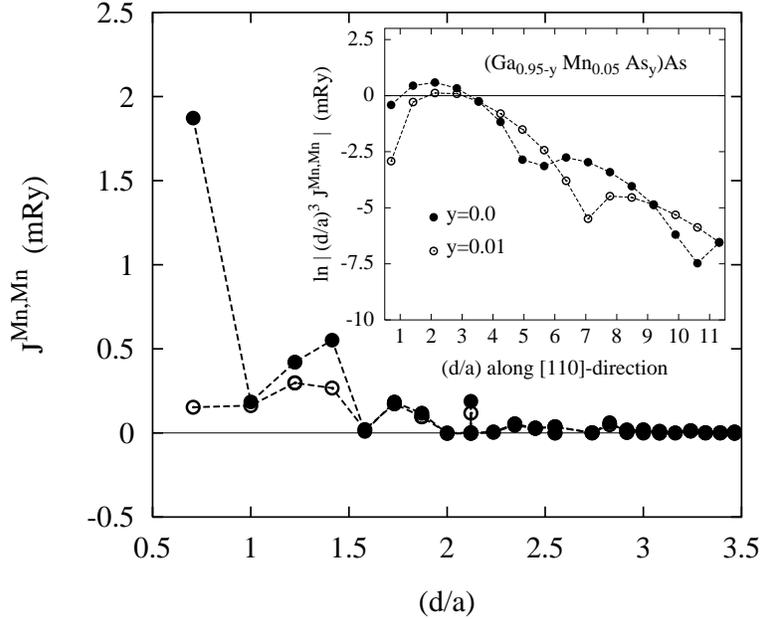}
\caption{Exchange interactions $J^{\rm Mn,Mn}$ between pairs of
Mn atoms in (Ga$_{0.95-y}$,Mn$_{0.05}$, As$_{y}$)As alloy plotted
as a function of their interatomic distance d (in units of the
lattice constant a).
Full and empty symbols correspond to $y=0$ and $y=0.01$,
respectively, where $y$ is the concentration of As-antisites
on the Ga-sublattice.
In the inset we show ${\rm ln} | (d/a)^{3} J^{\rm Mn,Mn}(d)|$ as
a function of the interatomic distance d along the [110]-direction
with dominating values of exchange interactions.}
\label{Fig.2}
\end{figure}

The present approach neglects the effect of local environment effects:
while the individual $J^{M,M}_{\bf R,R'}$ can be very different for
a particular environment, the corresponding configurationally averaged
${\bar J}^{M,M}_{\bf R,R'}$ in random systems are close to the CPA
value, Eq.~(\ref{eq_Jij}), as demonstrated recently \cite{nife}.
The CPA also correctly describes the concentration trends and the
carrier concentrations which, in turn, determine the size of the
alloy Fermi surface and thus the periods of oscillations.
The exchange interactions in GaMnAs alloys are exponentially damped 
due to the alloy disorder and by their halfmetallic character as
illustrated in Fig.~\ref{Fig.2} below.
The exchange interactions are also strongly anisotropic in the real
space due to the underlying zinc-blende lattice while an additional, 
but weaker anisotropy, could be due to the neglected spin-orbit effects.
Finally, because the hybridization between Mn- and host-atoms is included
by construction to all orders, both the RKKY-like and superexchange
interactions are included in the theory although their separation is
not possible in a simple way.

The typical results for exchange interactions in (Ga,Mn)As alloys are
illustrated in Fig.~\ref{Fig.2}.
We observe well pronounced ferromagnetic behavior of GaAs semiconductor
doped with 5\% of Mn-impurities: exchange interactions between
pairs of Mn atoms, $J^{\rm Mn,Mn}$, are ferromagnetic over number of
nearest neighbors with the dominating first nearest-neighbor interactions
$J^{\rm Mn,Mn}_{1}$ but without standard RKKY-oscillations.
Interactions are strongly reduced by the presence of native As-antisite 
defects, which reduce the number of free carriers: while each Mn-impurity 
introduces one hole into the host valence band, the As-antisite 
(and similarly, the Mn-interstitial) introduces two electrons.
With increasing concentration of As-antisites the leading
$J^{\rm Mn,Mn}_{1}$ becomes negative (antiferromagnetic) and 
such frustration coincides with the extinction of the ferromagnetism 
in the system (see below).
The damping of exchange interactions due to alloy disorder and, most
importantly, due to halfmetallic behavior of GaMnAs alloy, is illustrated
in the inset of Fig.~\ref{Fig.2}: the linear decrease of amplitudes
of the logarithm of exchange interactions multiplied by the RKKY-like
factor (${\rm ln} | (d/a)^{3} J^{\rm Mn,Mn}|$), indicating their 
exponential damping for large interatomic distances, is obvious.

\subsection{\label{TC} Curie temperatures of diluted (Ga,Mn)As alloys}

Here we briefly describe new approach for a quantitative determination
of the Curie temperature T$_{c}$ which correctly takes into account
the randomness in positions of magnetic impurities as well the
presence of native defects, like As-antisites and Mn-interstitials
and thus allows to explain their role in real samples.
The novel feature as compared to the conventional magnet thermodynamics
is the dilution, i.e., the presence of randomly distributed magnetic
defects of low, but finite concentration leading to the effect of
magnetic percolation.
It was clearly demonstrated recently \cite{GB2003} that a sophisticated
treatment of spin fluctuations using the RPA or Monte-Carlo methods but
without inclusion of disorder is itself unable to explain experimentally 
observed Curie temperatures in the framework of the parameter-free approach:
calculated critical temperatures are too high as compared to the experiment.
Recently, few groups have formulated parameter-free approaches which take
into account properly both disorder and spin-fluctuations and arrived at
a fair agreement with experiment for both well-annealed 
\cite{lars04,sato,EPL,schilff05,lars06} and as-grown samples \cite{GB05}.

The present accurate semi-analytical method separates the exact
treatment of disorder and the RPA treatment of spin-fluctuations 
and has three stages: (i) the first-principle determination of 
exchange parameters of random classical Heisenberg model \cite{our}; 
(ii) generation of a sequence of random configurations on a
disordered lattice (fcc Ga-sublattice for (Ga,Mn)As alloys) by
MC sampling technique; and (iii) for each configuration the
random Heisenberg model is treated analytically within the RPA.
As the lattice is random, the equations have to be solved numerically.

In the third step, the Green function $G_{\bf R,R'}$ for impurity
spins at sites $\bf R$ and $\bf R'$ satisfies
\begin{eqnarray}
(E - h^{\rm eff}_{\bf R}) \, G_{\bf R,R'}(E) &=& 2 \langle e^{z}_{\bf R} 
\rangle \delta_{\bf R,R'} - \langle e^{z}_{\bf R} 
\rangle \sum_{\bf R''} \, J_{\bf R,R''} \, G_{\bf R'',R'}(E) \, ,
\nonumber \\
h^{\rm eff}_{\bf R} &=& \sum_{\bf R''} J_{\bf R,R''} 
\langle e^{z}_{\bf R} \rangle \, .
\label{lrpa}
\end{eqnarray}
The quantity $h^{\rm eff}_{\bf R}$ is the local effective field acting
on the spin at the site $\bf R$, and $\langle e^{z}_{\bf R} \rangle$ 
is the local magnetic moment at the site $\bf R$ normalized with respect
to the magnetization averaged over all impurities. 
For a given temperature, $G_{\bf R,R'}(E)$ for impurity spins are
determined following the procedure similar to that of Callen \cite{callen}
resulting in the selfconsistent solution of the following set of equations
for the Curie temperature (for more details see paper \cite{GB05})
\begin{eqnarray}
k_{B} \, T_{c} = \frac{1}{3 N_{\rm imp}} \, 
\sum_{\bf R} \frac{1}{F_{\bf R}} \, ,
\;\;\;\;\;\;\; F_{\bf R} = \int_{-\infty}^{\infty} 
\frac{A_{\bf R,R}(E)}{E} \, dE \, ,
\nonumber \\
A_{\bf R,R}(E) = - \frac{1}{2 \pi} {\rm Im} \frac{G_{\bf R,R}(E)}
{\lambda_{\bf R}} \, ,
\;\;\;\; \lambda_{\bf R} = {\rm lim}_{T \rightarrow T_{c}} \, 
\langle e^{z}_{\bf R} \rangle \,/ {\bar e^{z}} \, ,
\label{tc}
\end{eqnarray}
where $N_{\rm imp}$ is the number of impurity sites and $\bar e^{z}$
denotes the averaged value of $e^{z}_{\bf R}$ over all impurity
sites.
At the end, the average over typically few hundred configurations
of the system, each of them including a few hundredth of impurity sites 
were enough to give robust values of the Curie temperature (with 
a numerical accuracy of about few K).
On the contrary, the MFA value as evaluated of the average lattice
which neglects the randomness in positions of magnetic ions is simply
$k_{B} T_{c}^{\rm MFA} = (2 x/3) \sum_{\bf R} J_{\bf 0,R}$, where
$x$ denotes the concentration of magnetic-atoms \cite{our}.

\begin{figure}
\centering
\includegraphics[width=0.75\textwidth]{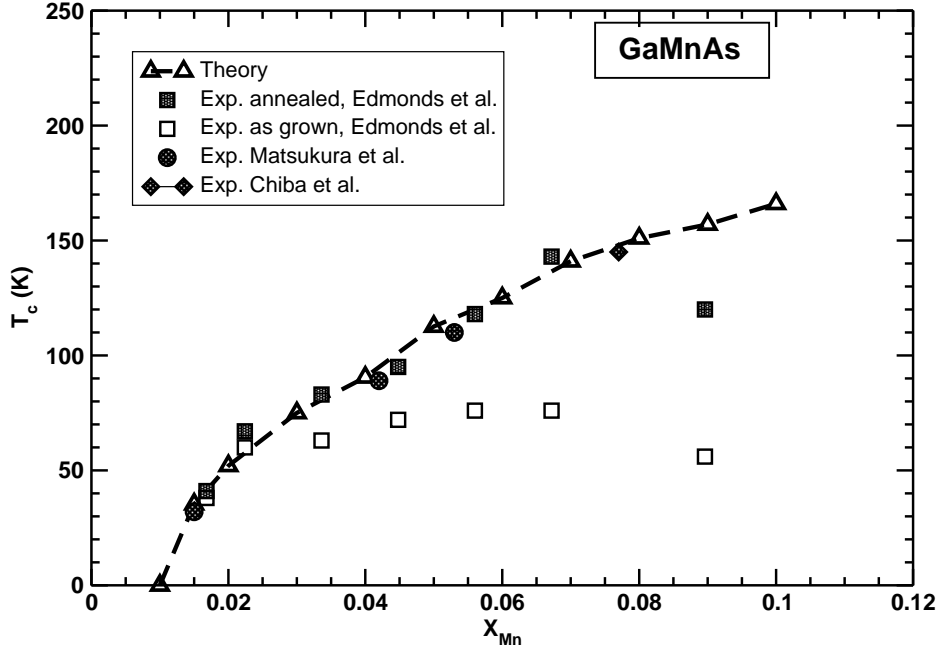}
\caption{Calculated Curie temperatures for (Ga$_{1-x}$ Mn$_{x}$)As 
alloys as a function of Mn-concentration. Theoretical results are
compared with available experimental data for both well annealed 
and as-grown samples.}
\label{Fig.3}
\end{figure}

The calculated T$_{c}$ according Eqs.~(\ref{tc}) are shown in
Fig.~{\ref{Fig.3} for the case of fully-annealed (Ga,Mn)As samples
without native defects (uncompensated samples) and with essentially 
random distribution of Mn-atoms \cite{EPL}.
In the same Figure we also show experimental results of 
Edmonds~{\it et al.} \cite{edmonds}, Matsukura \cite{matsukura}, 
and Chiba \cite{chiba}.
The agreement between theory and experiment is very good, except
for the single highest concentration (9~\%).
Our calculations thus suggests that for this specific concentration
the annealing is complete.
The present results are in a good agreement with related studies
\cite{lars04,lars06} based on Monte-Carlo simulations used
for the treatment of both spin-fluctuations and disorder: the
use of the same exchange parameters gives for 5~\% 137~K as
compared to the value of 125~K in the present approach.
Related approach which is based on the Korringa-Kohn-Rostoker
(KKR) Green functions developed in \cite{sato} gives 103~K.
The present theory also correctly predicts an expected threshold 
(about 1.5~\%) below which there is no ferromagnetism.
It should be noted that the theoretical threshold for the
occurrence of magnetism on the fcc lattice and the first 
nearest-neighbor Heisenberg model (19~\%) was also confirmed by 
numerical studies
\cite{sato,lars06}.
The calculated realistic exchange interactions, however, 
extends over several neighbor shells which reduces the effect
of magnetic percolation in realistic alloys.
On the other hand for as-grown samples \cite{edmonds,Ohno} the
Curie temperature is significantly reduced by native defects.

\begin{figure}
\centering
\includegraphics[width=0.75\textwidth]{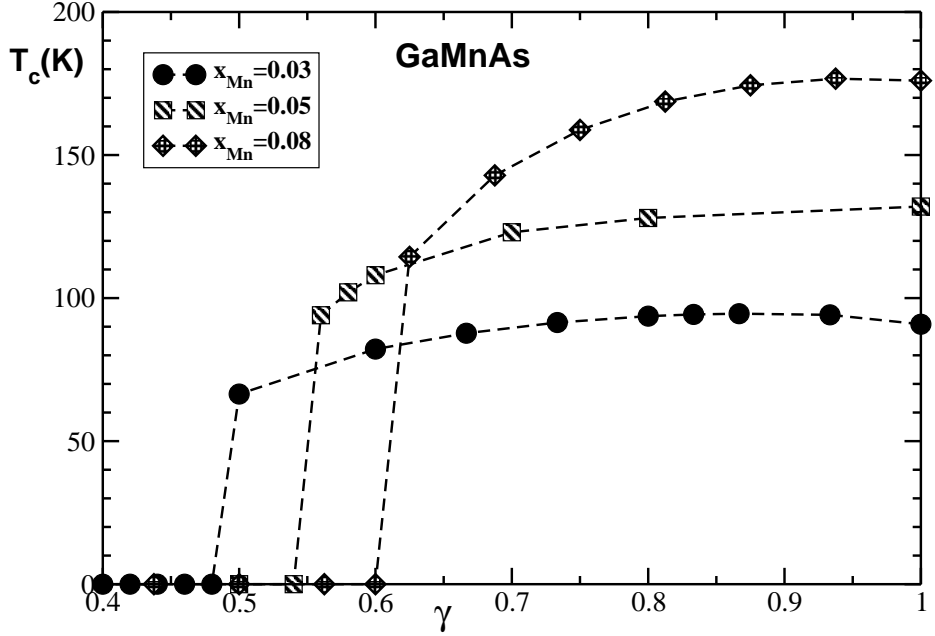}
\caption{The dependence of calculated Curie temperatures for 
(Ga$_{1-x-y}$ Mn$_{x}$ As$_{y}$)As alloys for three different
Mn-concentrations and for varying concentrations $y$ of 
As-antisites on Ga-sublattice plotted as a function of the 
carrier density. The parameter $\gamma$ is the ratio of the 
carrier concentration $n_{h}=x-2y$ and Mn-concentration $x$.}
\label{Fig.4}
\end{figure}

An immediate question is what is the proportion of two such
defects, Mn-interstitials and As-antisites in as-grown samples.
In order to answer this question we have first investigated
the effect of As-antisites.
The results are presented in Fig.~\ref{Fig.4}. clearly showing
a weak dependence of Curie temperatures on the effective
carrier concentration $\gamma$.
First, we observe ferromagnetic instabilities for smaller
values of $\gamma$. 
The reason for such instabilities is that for decreasing carrier 
concentrations are exchange interactions increasingly dominated 
by the antiferromagnetic (superexchange) contribution leading
to a frustration.
The experiment \cite{edmonds} for as-grown samples with a nominal 
Mn-concentration of $x=0.067$ shows, however, no ferromagnetic 
instability.
Second, relatively very weak dependence of calculated T$_{c}$
on $\gamma$ contradicts experiment which, on the contrary,
shows a pronounced dependence on the carrier concentration
\cite{edmonds}.
These facts seem to exclude As-antisites as a dominating 
mechanism for compensation in as-grown samples.

We will now demonstrate that observed Curie temperatures of
unannealed or compensated samples can be explained assuming 
that interstitials defects dominate \cite{GB05}. 
Such a dominance agrees with experimental observation of
Wolos {\it et al.} \cite{wolos} and Wang \cite{wang}.
We emphasize that a clear proof of the role of interstitials
is still necessary as other techniques like transmission
electron micrography \cite{anti1} or infrared absorption and
positron annihilation \cite{anti2} seem to indicate a much 
higher concentration of antisites.
On the other hand, recent theoretical study of Wu \cite{wu} also
seems to support the model of as-grown alloys described below.
The relevant quantity is the compensation parameter 
$\gamma=n_{h}/x_{\rm Mn}$, where $n_{h}$ is the carrier density 
and $x_{\rm Mn}$ is the nominal concentration of Mn-atoms. 
It should be noted that both $\gamma$ and $x_{\rm Mn}$ are
available from the experiment \cite{edmonds}.

Recent first-principle calculations \cite{masek} and
Rutherford backscattering experiments \cite{expinterst}
indicate that Mn-interstitials (Mn$_{I}$) are preferably 
attracted Mn-substitutional atoms on Ga-sublattice 
(Mn$_{\rm Ga}$) forming pairs of spins with a strongly
antiferromagnetic coupling.
We assume that Mn$_{I}$ are not completely random but
form, with a high accuracy, bound singlet pairs whose
effect on magnetically active ions is very small.
The remaining active Mn-atoms with effective concentration
$x_{\rm eff}=x_{\rm Mn(Ga)}-2 x_{\rm Mn(I)}$ which are not 
directly coupled to interstitials are assumed to be distributed
randomly and interact via the effective exchange coupling
corresponding to a measured carrier density $n_{h}$ (or,
alternatively via the experimentally determined parameter
$\gamma_{\rm eff}$).
The main features of the above model are also supported by 
recent first-principles calculations \cite{wu}.
Technically, one needs to determine exchange parameters,
Eq.~(\ref{eq_Jij}), but for concentration $x$ and the
effective number of carriers $n_{h}$ treated as
independent parameters.
This is not a straightforward task in the framework of the
first-principle theory without adjustable parameters,
but it could be achieved with help of co-doping by
impurities which have negligible influence on electronic
properties of the alloy at the Fermi energy which is relevant
for coupling between impurities \cite{our}.
For example, the Zn-doping on Ga-sublattice doping can increase
the number of carriers.
Similarly, the doping by Se-atoms on As-sublattice or 
by As-antisites decrease the number of carriers.
It should be noted that co-doping is used here as a purely 
calculational device to control carrier density while keeping 
calculations fully selfconsistent.

\begin{figure}
\centering
\includegraphics[width=0.75\textwidth]{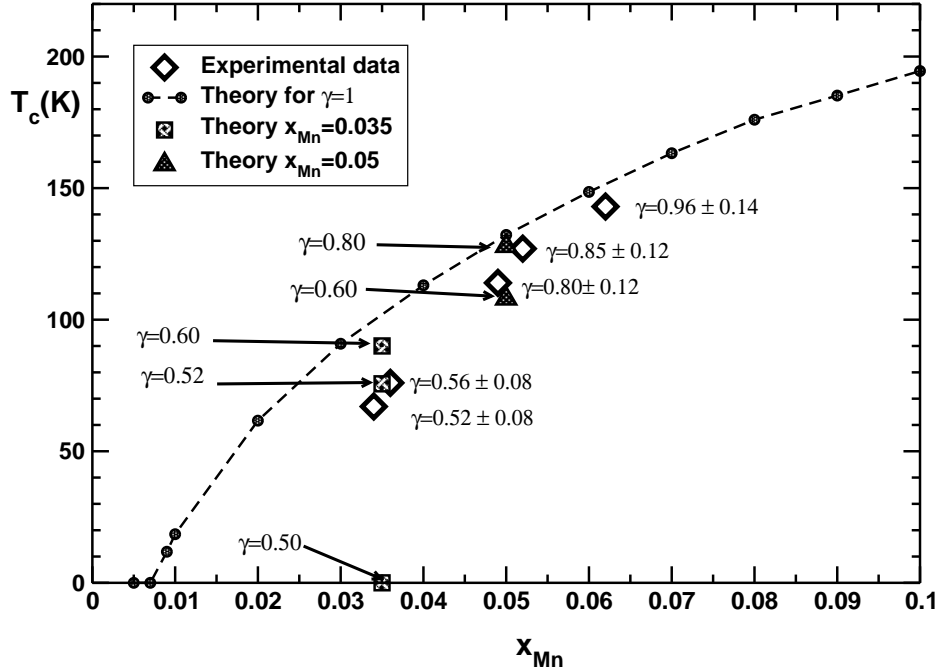}
\caption{Curie temperatures of GaMnAs alloys as a function
of Mn-concentration. Note that experimental data (diamonds)
are plotted for effective Mn-concentration of magnetically
active atoms and corresponding effective $\gamma$ as found
in experiment. Squares and triangles are Curie temperatures 
corresponding to Mn-concentrations 0.035 and 0.05, respectively, 
calculated for densities of holes that correspond to the experiment. 
The small circles (dashed line) correspond to uncompensated 
samples $\gamma=1$.}
\label{Fig.5}
\end{figure}

The results are summarized in Fig.~\ref{Fig.5} where,
for comparison, also results for $\gamma=1$ (see also
Fig.~{\ref{Fig.3}) are shown (small differences are
due to an improved statistics).
In the same figure are also given experimental data 
for nominal Mn-concentration of 0.067~\% but for a
different annealing corresponding to different
effective Mn-concentrations \cite{GB05}.
We observe, that well-annealed samples of highest T$_{c}$
are in very good agreement with the calculated values for
uncompensated samples ($\gamma=1$).
We remark that $\gamma=1$ curve can be accurately 
parameterized up to $x_{\rm Mn}=0.1$ by the curve
T$_{c} \approx A(x_{\rm Mn} - x_{c})^{1/2}$, where 
$x_{c}=0.0088$ and $A=649$~K.
Deviation from $\gamma=1$ curve is small for intermediate
T$_{c}$ but it becomes increasingly visible for
as-grown samples.
For example, T$_{c}$ of as-grown sample which corresponds
to $\gamma_{\rm eff}=0.52$  and $x_{\rm eff}=0.035$
agrees well with the calculated value (square symbol).
A similarly good agreement is obtained also for other,
differently annealed samples.
One can thus conclude that the present parameter-free 
theory is able to account for both well-annealed and
as-grown (Ga,Mn)As samples.
Based on a good agreement with experiment one can also 
conclude that the dominating mechanism for the reduction 
of the Curie temperature of as-grown samples as compared 
to well-annealed ones is the presence of Mn-interstitials.

\section{\label{CONC} Conclusions and prospects} 
We have presented a unified approach to describe electronic, 
structural, and magnetic properties of diluted magnetic semiconductors 
based on the GaAs host.
Our approach is based on the first-principles electronic structure
calculations which take into account low concentrations of various
defects present in the system as well as the finite lifetime due to
such defects.
The calculated total energies are then used to construct simple
effective Hamiltonians, namely the Ising model and the (classical) 
Heisenberg model, with parameters which are obtained from 
parameter-free first principles calculations.

We have shown that a combination of first principles electronic
structure calculations with a relatively simple methods
of statistical mechanics applied to model Hamiltonians can give 
a coherent picture of the phase stability and possible ordering 
as well as of the critical temperatures in diluted magnetic 
III-V semiconductors prepared under different conditions.
We have clearly demonstrated that inclusion of the randomness in 
positions of Mn-impurities is relevant for a good quantitative 
agreement between theory and experiment for both as-grown and 
well-annealed samples.
Based on a good agreement with experiment one can conclude that
the dominating mechanism for the reduction of the Curie temperature
of as-grown samples as compared to well-annealed ones is the presence
of Mn-interstitials.

One has to be aware that the samples studied in experiment need
not be in thermodynamical equilibrium, but rather are in a 
metastable state corresponding to a local minimum of the 
thermodynamical potential.
Phase stability studies based on equilibrium thermodynamics can 
nevertheless bring valuable information on certain trends in the 
structural evolution of DMSs and with respect to their stability,
including the magnetic one.

The basic information obtained from structural studies based on the
effective Ising model, e.g., a tendency to clustering, calculated
short-range order parameters, etc. can be in turn employed in the 
preparation of better structural model for sampling of magnetic 
impurities in the Heisenberg model as it was demonstrated recently 
\cite{satoim}.
We wish to mention that the same electronic structure model 
can be used to calculate the transport properties of DMSs, 
both in the bulk phase and for the multilayer arrangement, which 
are based on the Kubo-Greenwood and Kubo-Landauer approaches as
formulated in the framework of the CPA, respectively \cite{carva}.
In this way, the finite-lifetime effects due to various impurities 
present in the sample including native defects and possible co-dopants
are included consistently in the studies of structural, magnetic, and
transport properties of the DMS because they are based on the same 
electronic structure calculated from first principles.

\vspace{0.5cm}
\noindent {\small  Acknowledgments:} J.K. and V.D. acknowledge 
the financial support from the COST P19-OC150 project and from the
Grant Agency of the Academy of Sciences of the Czech Republic
(project A100100616).

\end{document}